# Non-uniqueness of second quantization of electromagnetic field: Stationary States of Jaynes-Cummings model in free space

A. Zh. Muradyan and G. A. Muradyan

*Department of Physics, Yerevan State University, 3750025, Yerevan, Republic of Armenia*

On the example of stationary states of a system consisting of an atom and a quantized electromagnetic field (the Jaynes-Cummings model in free space), it is shown that the physical characteristics of the system (as the energy and the probability distribution of finding the center of mass of the atom) explicitly depend on the choice of the basis of quantization of the electromagnetic field. Therefore, the secondary quantization procedure is not independent of the choice of the quantization basis in the interaction of the field with the material medium.

## I. INTRODUCTION

Second quantization or occupation number representation emerges from energy consideration of the free field and is regarded as basis independent. This means that at any choice of basis modes the field energy is presented in the diagonal form. In such a way it becomes possible to introduce the operators of creation and annihilation obeying certain commutation relations [1]. But, as shown in the next paragraph, another important quantity, the momentum of the field, does not necessarily have this property of basic independence: The diagonal form is obtained only by choosing the basis of the traveling waves. This is somewhat strange situation, but in itself has no internal contradiction or drawback. The quantum theory for such situation suggests introducing into consideration the measurement process. That is, it requires turning to the interaction, the quantities that can be measured in the experiment.

In this article, the problem is considered for a fundamental model in quantum optics, a two-level atom in the monochromatic field of electromagnetic radiation. It give the opportunity to present the physical content of the problem in a simple mathematical modeling.

The article is organized as follows. In Sec. II, we briefly repeat the conventional approach [2] of representing the electromagnetic field by modes. However, restricting ourselves to the one-dimensional case, we choose two possible modes in the form of an arbitrary superposition of traveling waves. Then, on this basis, we derive the expressions of the energy and momentum of the electromagnetic wave. It is shown that in the general case, in contrast to the expression for the field energy, the expression of the field momentum has a non-diagonal part. Further, in Sec. III, we introduce the boson operators of annihilation and creation, attributing them to the accepted bases of the general form. Based on this, we write down the expression of the potential energy of interaction of a quantized field with a two-level atom in a free space and compose the corresponding Hamiltonian and the total momentum of the system under consideration. The system description is also supplemented by the operator of the number of excitations in the system and by some operator of a combined nature. All four together make up a complete system of mutually commuting operators describing non-degenerate stationary states of the "atom+quantized electromagnetic field" system. The equations for determining the eigenfunctions and eigenvalues of the system are derived and briefly analyzed for dependence from the choice of the secondary quantization basis in Sec. IV. Finally, the eigenstates as well as the density matrix reduced to the translational motion of a two-level atom for a solvable system consisting of one photon and one atom are obtained and discussed in detail in Sec. V.

II. REPRESENTATION OF FREE ELECTROMAGNETIC FIELD ON AN ARBITRARY BASIS

Write Maxwell's curl equations in free space:

$$\nabla \times \boldsymbol{E} = -\frac{\partial \boldsymbol{B}}{\partial t}, \quad \nabla \times \boldsymbol{B} = \varepsilon_0 \mu_0 \frac{\partial \boldsymbol{E}}{\partial t}, \tag{1}$$

where $\varepsilon_0$ and $\mu_0$ are the free space permittivity and permeability, respectively, and $\varepsilon_0 \mu_0 = 1/c^2$.

First, consider the electric field. It follows from (1) together with $\nabla \cdot \boldsymbol{E} = 0$ the wave equation

$$\frac{\partial^2 \boldsymbol{E}}{\partial z^2} - \frac{1}{c^2}\frac{\partial^2 \boldsymbol{E}}{\partial t^2} = 0, \tag{2}$$

written in case of 1D problem. For a monochromatic wave of frequency $\omega$

$$\boldsymbol{E}(z,t) = \boldsymbol{E}(z)e^{-i\omega t} + \boldsymbol{E}^*(z)e^{i\omega t}. \tag{3}$$

In the following, $\boldsymbol{E}(z) = \mathbf{e}\, E(z)$, $\mathbf{e}$ being the unit vector of electric wave polarization and $E(z)$-the complex amplitude of the field. This, according to (2) and (3), satisfies the equation

$$\frac{d^2 E(z)}{dz^2} + \frac{\omega^2}{c^2} E(z) = 0 \tag{4}$$

with simple linearly independent solutions $e^{ikz}$ and $e^{-ikz}$, $k = \omega/c$. In this paper, however, we do not choose them as basis states, but arbitrary linear combinations

$$\chi_1(z) = \cos\alpha\, e^{ikz} + \sin\alpha\, e^{-ikz}, \quad \chi_2(z) = -\sin\alpha\, e^{ikz} + \cos\alpha\, e^{-ikz}, \tag{5}$$

where $\alpha$ is an arbitrary real number. Now the solution of equation (4) will be written as

$$E(z) = f_1 \chi_1(z) + f_2 \chi_2(z)$$

decomposition. Accordingly, we will have

$$\boldsymbol{E}(z,t) = \mathbf{e}\left((f_1 \chi_1(z) + f_2 \chi_2(z))e^{-i\omega t} + c.c.\right). \tag{6}$$

Other field component, magnetic field induction of the monochromatic wave is represented as

$$B(z,t) = B(z)e^{-i\omega t} + B^*(z)e^{i\omega t}, \qquad (7)$$

the vector amplitude $B(z)$ of which is determined, according to the first one of Eq. (1), through the amplitude $E(z)$ of the electric field by the equation

$$B(z) = \frac{1}{i\omega} \nabla \times E(z).$$

Substitution of $E(z) = e\, E(z)$, $\nabla = e_z \frac{d}{dz}$ and explicit expression of bases from the Eq. (5) yields

$$\begin{aligned} B(z) &= \frac{1}{i\omega}(e \times e_z)\frac{dE(z)}{dz} \\ &= \frac{1}{c}(e \times e_z)\left(-f_1\left(\cos\alpha\, e^{ikz} - \sin\alpha\, e^{-ikz}\right) + f_2\left(\cos\alpha\, e^{-ikz} + \sin\alpha\, e^{ikz}\right)\right). \end{aligned} \qquad (8)$$

Having expressions of electric and magnetic fields, we proceed to the expressions of energy

$$H_{field} = \frac{1}{2}\varepsilon_0 E^2(z,t) + \frac{1}{2\mu_0}B^2(z,t) = \varepsilon_0 E^*(z)\cdot E(z) + \frac{1}{\mu_0}B^*(z)\cdot B(z)$$

and the momentum

$$P_{field} = \frac{1}{c^2}S(z,t) = \varepsilon_0 E(z,t) \times B(z,t) = \varepsilon_0\left(E(z) \times B^*(z) + E^*(z) \times B(z)\right)$$

of the field, were $S(z,t)$ is the Pointing vector. The corresponding substitutions and some algebra lead to

$$H_{field} = 2\varepsilon_0\left(f_1^* f_1 + f_2^* f_2\right), \qquad (9)$$

$$P_{field} = e_z \frac{2\varepsilon_0}{c}\left(\cos 2\alpha\left(f_1^* f_1 - f_2^* f_2\right) - \sin 2\alpha\left(f_1^* f_2 + f_2^* f_1\right)\right). \qquad (10)$$

A remarkable new point here is the second term in the momentum expression of the field. In general, it is not diagonal with respect to the basis states (indices), and becomes diagonal only if $\alpha = 0$, that is, for the basis of the counter-propagating waves.

## III. FIELD QUANTIZATION AND COMPLETE SET OF OPERATORS FOR ATOM+QUINTIZED FIELD SYSTEM

The electromagnetic field is quantized by attributing to $f_{1,2}$ and $f_{1,2}^*$ boson annihilation $\hat{a}_{1,2}$ and creation $\hat{a}_{1,2}^+$ operators respectively. For this purpose, the Born-von Karman periodicity condition $E(z+L) = E(z)$ is traditionally set, where $L$ is an arbitrary length playing an intermediate role in the calculation of the values measured in the experiment.

$$f_{1,2} \to \sqrt{\frac{\hbar\omega}{2L\varepsilon_0}}\,\hat{a}_{1,2}, \quad f_{1,2}^{*} \to \sqrt{\frac{\hbar\omega}{2L\varepsilon_0}}\,\hat{a}_{1,2}^{+}.$$

In terms of $\hat{a}_{1,2}$ and $\hat{a}_{1,2}^{+}$, the Hamiltonian (9) and the momentum (10) yield

$$\hat{H}_{field} = \hbar\omega\left(\hat{a}_1^{+}\hat{a}_1 + \hat{a}_2^{+}\hat{a}_2\right), \tag{11}$$

$$\hat{P}_{field} = \mathbf{e}_\mathbf{z}\frac{\hbar\omega}{c}\left(\cos 2\alpha\left(\hat{a}_1^{+}\hat{a}_1 - \hat{a}_2^{+}\hat{a}_2\right) - \sin 2\alpha\left(\hat{a}_1^{+}\hat{a}_2 + \hat{a}_2^{+}\hat{a}_1\right)\right). \tag{12}$$

respectively. As expected, they are commutative.

The electric field intensity of the monochromatic wave appears as

$$\hat{\mathbf{E}}(z,t) = \mathbf{e}\, E_0\left(\left(\left(\cos\alpha\, e^{ikz} + \sin\alpha\, e^{-ikz}\right)\hat{a}_1 + \left(\cos\alpha\, e^{-ikz} - \sin\alpha\, e^{ikz}\right)\hat{a}_2\right)e^{-i\omega t} + H.c.\right), \tag{13}$$

where $E_0 = \sqrt{\hbar\omega/2L\varepsilon_0}$ and *H.c.* stands for Hermitian conjugate.

It is well known that the interaction of the classical radiation field $\mathbf{E}(z,t)$ with a two level atom is presented by the form $-\hat{\mathbf{d}}\cdot\mathbf{E}(z,t)$, where $\hat{\mathbf{d}}$ is atomic dipole moment operator. By quantization of the field, the interaction energy takes the following form in the rotating wave approximation:

$$\hat{H}_{inter} = -\beta\hat{\sigma}^{+}\left(\left(\cos\alpha\, e^{ikz} + \sin\alpha\, e^{-ikz}\right)\hat{a}_1 + \left(\cos\alpha\, e^{-ikz} - \sin\alpha\, e^{ikz}\right)\hat{a}_2\right) - H.c., \tag{14}$$

where $\beta = dE_0$, $\hat{\sigma}^{+} = \begin{pmatrix} 0 & 1 \\ 0 & 0 \end{pmatrix}$ is the operator of rising the population of the atom in the matrix representation.

We also add the Hamiltonian of the atom, including there the internal two-level state and the translational motion of the center of mass:

$$\hat{H}_{atom} = -\frac{\hbar^2}{2M}\frac{d^2}{dz^2} + \frac{\hbar\omega_0}{2}\left(1 + \hat{\sigma}_3\right), \tag{15}$$

Thus we come to the expressions of the Hamiltonian and momentum for the complete "two level atom+quantized field" system:

$$\hat{H} = \hat{H}_{atom} + \hat{H}_{field} + \hat{H}_{inter}, \tag{16}$$

$$\hat{P} = \hat{P}_{atom} + \hat{P}_{field}, \tag{17}$$

where $\hat{P}_{atom} = (-i/\hbar)d/dz$ is atomic momentum operator.

Two more operators describing the states of the system are known [3]: the operator of the number of excitations

$$\hat{N} = \frac{1}{2}(1 + \hat{\sigma}_3) + \hat{a}_1^{+}\hat{a}_1 + \hat{a}_2^{+}\hat{a}_2 \tag{18}$$

and the combined operator

$$\hat{T} = \hat{\sigma}_3 \exp\left(\frac{i\pi \hat{P}_{atom}}{\hbar k}\right) = \hat{\sigma}_3 \exp\left(\frac{\pi}{k}\frac{d}{dz}\right). \quad (19)$$

These four operators, $\hat{H}$, $\hat{P}$, $\hat{N}$ and $T$, constitute a complete set of mutually commutative operators that determine the nondegenerate states of the system under consideration, which has four degrees of freedom: the translational and internal degrees of freedom of the atom and two modes of the quantized field.

## IV. DERIVATION OF EQUATIONS FOR STATIONARY STATES

An important feature of our quantum model is that it is in free from material bodies (mirrors) space. It is actually an extended version of the well-known Jaynes-Cummings model [4, 2].

We are looking for state vectors that are simultaneously eigenvectors of all four of the above mentioned operators. Note that since all four operators are independent of time, their non-degenerate eigenstates will also be stationary states. They can be written in the following general form:

$$|\Psi(z)\rangle = \begin{pmatrix}0\\1\end{pmatrix}\sum_{m=0}^{N} a_m(z)|m\rangle_1|N-m\rangle_2 + \begin{pmatrix}1\\0\end{pmatrix}\sum_{m=0}^{N} b_m(z)|m\rangle_1|N-1-m\rangle_2, \quad (20)$$

where the column factors represent the atomic state alternately on the lower and upper energy levels. The Dirac notation $|m\rangle_{1,2}$ is used for photonic number state. $N$ is the number of excitations in the system, coinciding with the number of photons, if the atom is on the lower level, and one more than the number of photons, if the atom is on the upper level. The subjects of the search are probabilistic amplitudes $a_m(z)$ and $b_m(z)$, the first of which corresponds to the states with the atom on the lower energy level, and the second-on the excited energy level.

Expression (20) automatically (for any $a_m(z)$ and $b_m(z)$) satisfies the eigenvalue equation of the $\hat{N}$ operator: $\hat{N}|\Psi(z)\rangle = N|\Psi(z)\rangle$. If eigen equations are satisfied also for the Hamiltonian $\hat{H}$ and the total momentum $\hat{P}$, then the equation for the combined operator T is satisfied by itself. Therefore, the amplitudes $a_m(z)$ and $b_m(z)$ are determined by energy equation $\hat{H}|\Psi(z)\rangle = E|\Psi(z)\rangle$ and momentum equation $\hat{P}|\Psi(z)\rangle = P|\Psi(z)\rangle$. Substituting there Eqs. (11), (14)-(17) and (20) we come to the following closed system of differential-difference eigen equations:

$$\frac{d^2 a_m(\eta)}{d\eta^2} + (\varepsilon - N\hbar\Omega)a_m(\eta) =$$
$$-\varsigma\left(\cos\alpha\, e^{-i\eta} + \sin\alpha\, e^{i\eta}\right)\sqrt{m}\, b_{m-1}(\eta) - \varsigma\left(\cos\alpha\, e^{i\eta} - \sin\alpha\, e^{-i\eta}\right)\sqrt{N-m}\, b_m(\eta), \quad (21a)$$

$$\frac{d^2 b_m(\eta)}{d\eta^2} + (\varepsilon - N\hbar\Omega - \Delta) b_m(\eta) = \qquad (21b)$$
$$-\varsigma\left(\cos\alpha\, e^{i\eta} + \sin\alpha\, e^{-i\eta}\right)\sqrt{m+1}\, a_{m+1}(\eta) - \varsigma\left(\cos\alpha\, e^{-i\eta} - \sin\alpha\, e^{i\eta}\right)\sqrt{N-m}\, a_m(\eta),$$

$$-i\frac{da_m(\eta)}{d\eta} + \cos 2\alpha\,(2m-N) a_m(\eta) \qquad (22a)$$
$$-\sin 2\alpha\left(\sqrt{m(N-m+1)}\, a_{m-1}(\eta) + \sqrt{(m+1)(N-m)}\, a_{m+1}(\eta)\right) = p\, a_m(\eta),$$

$$-i\frac{db_m(\eta)}{d\eta} + \cos 2\alpha\,(2m-N+1) b_m(\eta) \qquad (22b)$$
$$-\sin 2\alpha\left(\sqrt{m(N-m)}\, b_{m-1}(\eta) + \sqrt{(m+1)(N-m-1)}\, b_{m+1}(\eta)\right) = p\, b_m(\eta),$$

$m = 0, 1, \ldots, N$. Eqs. (21a,b) come from the Hamiltonian equation, Eqs. (22a, b) come from the momentum equation. The equations are written in dimensionless quantities: $\eta = k z$, $\varepsilon = E/E_{rec}$, $\Omega = \omega/E_{rec}$, $\Delta = \hbar(\omega-\omega_0)/E_{rec}$, $\varsigma = \beta/E_{rec}$, and $E_{rec} = \hbar^2 k^2/2M$, where $M$ is the mass of the atom.

## V. EXPLICIT SOLUTIONS IN CASE OF $N = 1$

The subject of interest is the dependence of the solutions of the equations on the parameter $\alpha$, which determines the basis of second quantization. It turns out, firstly, that the choice of the basis of traveling waves ($\alpha = 0$) by introducing new amplitudes

$$\bar{a}_m(\eta) = a_m(\eta) e^{i 2m\eta}, \quad \bar{b}_m(\eta) = b_m(\eta) e^{i(2m+1)\eta}$$

frees the coefficients of the system (21a)-(22b) from the coordinate dependence. This directly means the discrete energy spectrum of the system (for a certain value of the momentum $p$). The situation is fundamentally different if another basis is chosen, for instance, the basis of standing waves: the energy spectrum is of band structure. And other observables differ qualitatively. To identify these inconsistencies in more detail, consider a special case $N = 1$. It can be studied analytically explicitly for the regime of large resonance detunings,

$$\Delta \gg \varepsilon - N\hbar\Omega + \frac{1}{b_m(\eta)}\frac{d^2 b_m(\eta)}{d\eta^2},$$

significantly exceeding the high-frequency shift of the energy levels of the atom under the influence of the photon field. Then $|b_m(\eta)|^2 \to 0$ and $a_m(\eta)$ satisfies the following recurrence equation:

$$\frac{d^2 a_{0,1}(\eta)}{d\eta^2} + \left(\varepsilon - \hbar\Omega + \xi(1 \mp \sin 2\alpha \cos 2\eta)\right) a_{0,1}(\eta) = -\xi\left(\cos 2\alpha \cos 2\eta \pm i \sin 2\eta\right) a_{1,0}(\eta), \quad (23a)$$

$$i\frac{da_{0,1}(\eta)}{d\eta} \pm \cos 2\alpha\, a_{0,1}(\eta) + p\, a_{0,1}(\eta) = -\sin 2\alpha\, a_{1,0}(\eta). \tag{23b}$$

The upper signs correspond to the first indexes, the lower-to the second indexes of the sought functions.

In the case of the basis of traveling waves, the system (23a,b) is easily solved. The energy spectrum has two branches and is given by the formula

$$\varepsilon_\pm = \hbar\Omega + 1 + p^2 - \xi \pm \sqrt{4p^2 + \xi^2}. \tag{24}$$

On both branches, the probability amplitudes are given by expressions

$$a_0(\eta) = a_0\, e^{i(p+1)\eta}, \quad a_1(\eta) = a_1\, e^{i(p-1)\eta}, \tag{25}$$

where $a_0$, $a_1$ are definite constants connected by the normalization condition $a_0^2 + a_1^2 = 1$.

In the case of the standing wave basis (at $\alpha = \pi/4$), the state amplitude $a_0(\eta)$ satisfies the following of (21a)-(22a) equation

$$\frac{d^2 a_0(\eta)}{d\eta^2} + \xi \sin 2\eta \frac{da_0(\eta)}{d\eta} + \left(\varepsilon - \hbar\Omega + \xi(1 - \cos 2\eta - i p \sin 2\eta)\right) a_0(\eta) = 0,$$

which for the related function $\bar{a}_0(\eta) = a_0(\eta)\exp(-\xi/4 \cdot \sin 2\eta)$ is converted into a Schrödinger form

$$\frac{d^2 \bar{a}_0(\eta)}{d\eta^2} + \left(\varepsilon - \hbar\Omega + \xi(1 - 2\cos 2\eta - i p \sin 2\eta) + \xi^2 \sin^2 2\eta\right)\bar{a}_0(\eta) = 0. \tag{26}$$

Eq. (26) is an equation with periodic coefficients, a Hill-type equation, and according to the Floquet-Bloch theory the eigenvalue spectrum (the energy spectrum) has a band structure.

Neglecting the term with a small coefficient $\xi^2$ in Eq. (26), we arrive at Mathieu's equation [5] with two linearly independent solutions

$$\bar{a}_{0,0}(\eta) = y_0(\varepsilon - \hbar\Omega + \xi, \sqrt{1 - p^2/4}\,\xi, \eta - i\phi), \quad \bar{a}_{0,1}(\eta) = y_1(\varepsilon - \hbar\Omega + \xi, \sqrt{1 - p^2/4}\,\xi, \eta - i\phi)$$

in the region $0 \leq \eta \leq \pi$ and periodically continuing according to the Bloch relation $\bar{a}_0(\eta) = \exp(i\pi p_{quasi})\bar{a}_0(\eta - \pi)$. Here $\phi = 1/2 \cdot \text{Arcch}\left(1/\sqrt{1 - p^2/4}\right)$, and the quasimomentum $p_{quasi}$ is connected with energy $\varepsilon$ by the familiar dispersion relation, forming the band structure of energy spectrum. Numerical calculations add expected result that the widths of forbidden zones are monotonically narrowing with increasing total momentum p.

The amplitude $a_1(\eta)$ is calculated by the relation

$$a_1(\eta) = \left(-i\frac{d}{d\eta} - p\right) a_0(\eta),$$

following from the first of Eqs. (23b).

Another quantity of direct interest for comparison at different selections of bases of field quantization is the spatial distribution of the center of mass of the atom, which is clear from the ideological side and the easiest to implement in the experiment. According to the quantum theory, this distribution is given by the element $\rho_{atom}$ of the reduced density matrix, which is obtained from the density matrix of the complete system $\hat{\rho} = |\Psi(z)\rangle\langle\Psi(z)|$ by averaging over the complete system of states $|m\rangle_1 |N-m\rangle_2$ of the photon subsystem. In the result of calculations we get

$$\rho_{atom} = |a_0(\eta)|^2 + |a_1(\eta)|^2. \qquad (27)$$

For the basis of standing waves, the density (27) periodically changes in space, while in the case of the basis of traveling waves does not change at all.

## VI. CONCLUSIONS

Describing the electromagnetic field and its interaction with the medium, we usually decompose it on a certain basis. In classical field theory, the experimentally measurable quantities does not depend on the choice of basis. This, in particular, applies to the bases of traveling waves and standing waves, which directly follows from the equivalence of Fourier expansions performed on these basis functions. This freedom of choice of the bases, as it seems to us, by inertia, is also attributed to the theory of the second quantization, of course, under the same external (boundary) conditions. The different boundary conditions by the bases of the traveling wave and the standing wave really create differences between the measured values of the system [6, 3].

Present approach considers boundary constraints, namely, the interaction with mirrors of an optical resonator [6], merely as a type of interaction. It follows from this a generalization that any interaction, for example with a two-level atom, should eliminate the equivalence of quantization bases inherent in free quantum fields. To test this idea, we chose the well-known model of James-Cummings, having considered it in free space. Arbitrary superposition of traveling waves is chosen as a basis for secondary quantization. Two "most different bases", the basis of traveling waves and the basis of standing waves, reveal fundamentally different results for the energy spectrum of the system and for the spatial distribution of the c.m. of the atom. The system has only two possible values of energy for a given momentum " p " and a homogeneous distribution of the detection probability of the atom in the space in the case of the basis of traveling waves, but the band structure of energies and a periodic spatial distribution of c.m. of atom in the case of basis of standing waves. All this leads us to the conclusion that the secondary quantization of the electromagnetic field in quantum optics is not a uniquely defined procedure. A reasonable way out of the situation can be the transition to the operators of the creation and annihilation of the photon on the basis of traveling waves, in accordance with the original ideas of Einstein and de Broglie.

Note that the first results on the discrepancy (in the time evolution of atom+quantized field system in free space) due to the various bases of the second quantization were announced at the conference [7].

----------------------------------------------------------------------------------------